\documentclass{article}%
\usepackage{amsfonts}
\usepackage{amssymb}
\usepackage{amsmath}
\usepackage{graphicx}%
\providecommand{\U}[1]{\protect\rule{.1in}{.1in}}
\pdfoutput=1
\begin{document}

\title{Why `Fair Market Valuations' are Inappropirate for Employee-Owned Firms and Partnerships}
\author{David Ellerman\\University of Ljubljana, Slovenia\\david@ellerman.org\\https://orcid.org/0000-0002-5718-618X}
\date{\begin{center}
		Forthcoming in:
		International Review of Applied Economics
\end{center}}

\maketitle

\begin{abstract}
\noindent The usual formulas for the fair market valuation of a firm at time
$t$ include the profits accruing to the shares at time $t$ from the use of
wage or salaried labor in the future. But in employee-owned firms or
partnerships, the future worker-members or partners are the residual claimants
at those future times, so in those cases, the future residuals do not accrue
to the current shareholder/residual-claimants. Hence any `fair market
valuation' of an employee-owned firm or partnership that assumes those future
residuals accrue to the current shareholder/residual-claimants is inappropriate.


\end{abstract}
\tableofcontents

\section{Introduction}

The basic argument of this paper can be easily outlined. The usual formulas
for market or so-called `fair' valuations of a firm are equivalent to the
formula that breaks into two parts: (1) the present property rights, the net
asset value of the firm, plus (2) the present value of the expected future
profits. This formula assumes that the rights of the current owners of the
firm include the residual or profit claims in the future time periods. But
this is not true in a partnership or in an employee-owned firm such as an
Employee Stock Ownership Plan (ESOP) (for simplicity, assume a 100\% ESOP) or
a worker cooperative. In both those cases, the future partners or the future
employee-owners are the residual/profit claimants in those future time
periods, so the present value of those profits cannot be imputed to the
current partners or employee-owners.

In other words, the future partners or employee-owners are not the employees
of the current ones. In a \textit{conventional} firm, the claim of future
workers on future revenues is capped at their wages or salaries, so the
expected amount of future residual profits (revenues minus expenses including
employee expenses) can be plausibly assigned to the current owners, at least
in terms of expectations. But partnerships, ESOPs, and worker cooperatives are
not conventional firms in that respect. The future partners or worker-members
are just as much residual claimants as the current partners or worker-members;
their claims are not capped at wages or salaries. Hence the residual profits in
those future time periods are imputed or assigned to those future residual
claimants, not the current partners or worker-members.

It is a simple argument. But the predominance of conventional firms leads to
the standard valuations being taken as ``objective'' by professional or certified valuators. Their
professional status as certified business valuators would be jeopardized by
any major deviation from the standard `objective' formulas.\footnote{One is
reminded of Upton Sinclair's famous quip: ``It is difficult to
get a man to understand something when his salary depends upon his not
understanding it,''} Moreover, the full arguments depend on
some of the mathematics of financial economics, e.g., by the Nobel laureates
Merton Miller and Franco Modigliani \cite{Miller-Modig:valuation}, that go far
beyond the cookbook formulas taught to business valuators. Our treatment in
the body of the paper will only use simple mathematics or heuristics with the
Miller-Modigliani methods treated in the Appendix.

\section{Fair market valuation of an asset}

The relevant points can be made using a simple model. Production assets like a
machine or building have passive or active uses. In the passive usage, the
asset is rented out for (we assume) some going market rental rate. In
competitive markets, the cost of the asset would be the discounted present
value of rentals (net of maintenance costs) plus the salvage value. In the
active use of the asset, the asset is combined with a set of complementary
production services (various inputs and labor) into a going-concern firm and
then the owner of the asset would value the asset at the discounted present
value of the future net returns plus the salvage value.

\subsection{The passive use of the asset}

Consider a capital asset with a market cost of $C$ which yields the annual
stream of capital services $K,...,K$ for $n$ years (for simplicity, assume no
depreciation) and then has a salvage value of $S$. The capital services can be
rented or leased out for $R$ per year and the interest rate is assumed to be
$r$ (e.g., $0.10$ for $10\%$). Hence the discounted present value of the
services rendered by the asset on the rental market is:

\begin{center}
$\sum_{k=1}^{n}\frac{RK}{(1+r)^{k}}+\frac{S}{(1+r)^{n}}=RKa(n,r)+\frac
{S}{(1+r)^{n}}$
\end{center}

\noindent where $a\left(  n,r\right)  $ is called the \textit{present value of
an ordinary annuity of one }\cite[p. 52]{friedman-ordway:appraisal}:

\begin{center}
$a\left(  n,r\right)  =\frac{1}{\left(  1+r\right)  }+\frac{1}{\left(
1+r\right)  ^{2}}+...+\frac{1}{\left(  1+r\right)  ^{n}}$.

Then competitive arbitrage between the buy and lease markets will enforce the equation:

$C=RKa(n,r)+\frac{S}{(1+r)^{n}}$.
\end{center}

\subsection{The active use of the asset}

In the active case, a supplementary set of inputs is purchased at the variable
cost $VC$ and $L$ units of labor are hired at the wage rate of $W$ per year,
and outputs $Q$ are produced which sell for unit price $P$. Then the economic
(i.e., pure) profits obtained at the end of each year are:

\begin{center}
$\pi=PQ-RK-VC-WL$
\end{center}

\noindent(where $RK$ is the implicit cost of tying up that asset in the
going-concern firm). Hence the discounted present value of the economic
profits is:

\begin{center}
$V_{0}:=\pi a\left(  n,r\right)  $.
\end{center}

The \textit{capitalized value of the capital asset} is defined as the
discounted present value of the income stream ``generated'' by the capital asset. But we have seen there are
two ways to generate an income stream from the asset, the passive and the
active ways. In the active case, the cashflow stream generated by the asset
is: $PQ-VC-WL$ per year plus the salvage value so the total discounted present
value in the active case is:

\begin{center}
$V:=\left(  PQ-VC-WL\right)  a\left(  n,r\right)  +\frac{S}{\left(
1+r\right)  ^{n}}=\left(  RK+\pi\right)  a\left(  n,r\right)  +\frac
{S}{\left(  1+r\right)  ^{n}}=\left(  RKa\left(  n,r\right)  +\frac{S}{\left(
1+r\right)  ^{n}}\right)  +\pi a\left(  n,r\right)  =C+V_{0}$.
\end{center}

This simple example illustrates the capitalized value of the capital asset
formula that is behind the valuation formulas for valuing the shares in a corporation.

\begin{quotation}
\noindent There, in valuing any specific machine, we discount at the market
rate of interest the stream of cash receipts generated by the machine, plus
any scrap or terminal value of the machine, and minus the stream of cash
outlays for direct labor, materials, repairs, and capital additions. The same
approach, of course, can also be applied to the firm as a whole, which may be
thought of in this context as simply a large, composite machine. \cite[p.
415]{Miller-Modig:valuation}
\end{quotation}

Thus we see that Miller and Modigliani use the same (active case) capitalized
value of the capital asset, i.e., the ``specific machine,'' since those net cash receipts are $\left(PQ-VC-WL\right)  $ and the terminal value is $S$ all discounted back to
present value at the interest rate to give $V$. That is a standard formula for
the capitalized value of a capital asset. As John Maynard Keynes put it:

\begin{quotation}
\noindent When a man buys an investment or capital-asset, he purchases the
right to the series of prospective returns, which he expects to obtain from
selling its output, after deducting the running expenses of obtaining that
output, during the life of the asset. \cite[p. 135]{keynes:GeneralTheory}
\end{quotation}

\subsection{Analysis of the standard valuation formulas}

The same formula is applied to a corporation as a ``  large,
composite machine.'' Then the net value of the
``machine'' $C$ is replaced by the
\textit{Net Asset Value} ($NAV$) of the corporation's assets and liabilities,
e.g., on the year-end balance sheet of the corporation. The discounted present
value of the future pure profits, i.e., $V_{0}$ in the model, is usually
called the \textit{Goodwill}. The usage of the word \textquotedblleft
goodwill'' to describe ``intangible
assets'' such as ``reputation'' do not add to the capitalized valuation of the
corporation unless they add to future profits. Hence the valuation formula
$V=C+V_{0}$ is the ``NAV + Goodwill'' version
of the formula for $V$.

The valuation formula $V=\left(  PQ-VC-WL\right)  a\left(  n,r\right)
+\frac{S}{\left(  1+r\right)  ^{n}}$ is the \textit{discounted future
cashflow} version of the valuation formula that is essentially used in
cookbook versions by practicing business valuators. Thus we have the
equivalence between two different ways to express the standard
``fair market valuation'' of a corporation in our simple model:

\begin{center}
Discounted Cashflow Valuation = Net Asset Value + Goodwill Valuation
\end{center}

The NAV + Goodwill formula is particularly important because it parses $V$
into the actual owned assets and owed liabilities of the corporation $NAV$
plus the anticipated or expected market contracts that the company
``expects to obtain'' in the future (see
\cite{ell:goodwill}; \cite[Chapter 3]{ell:juris-econ}). But the corporation
has no present property rights to force the customers to buy the outputs for
$PQ$ or the suppliers to supply the inputs and labor for $VC+WL$. Future
expectations, no matter how carefully extrapolated from past data, are not
present property rights. That seems to be the underlying reason why U.S. or
international accounting standards do not allow (``internally
generated\textquotedblright) goodwill to be listed on the company's balance
sheet as an owned asset.

This parsing of the ``fair market valuation'' between actual assets and liabilities $NAV$ plus anticipated or expected
future assets and liabilities $V_{0}$ is behind an important distinction in
the debates about corporate governance. If we abstract away all the future
market contracts with customers and suppliers from a corporation operating as
a going-concern (the active case), then we are left with the bare-corporation
holding the actual assets and liabilities with the NAV valuation (the passive
case). Some commentators have suggested the term ``firm'' to denote the corporation embedded in the market
relationships as opposed to the (bare) corporation itself, e.g.,
``the firm--the organization built via contracts transferring
control over resources to the corporations used to legally structure the
firm'' \cite[p. 4]{robe:firm}. This difference between the
(market-embedded) firm and the (bare) corporation is the anticipated future
contracts with customers and suppliers to which the corporation has no present
property rights. Hence there is no present ``ownership of the
\textit{firm}'' \cite{ell:ownership-myth}, only the ownership
of the (conventional bare) corporation by the current shareholders.

\section{Property rights versus personal rights}

In a conventional corporation, the owners (shareholders) are the claimants of
the residual after all liabilities are met, i.e., the \textquotedblleft
residual claimants.'' But other types of corporations such as
worker cooperatives, e.g., the Mondragon cooperatives \cite{whytes:mondragon},
have residual claimancy determined in another manner. To understand the
difference, we need to separate personal rights from property rights.

A \textit{personal right} is attached to a person because they qualify for it
by having a certain functional role. For instance, certain persons qualify for
citizenship in a country (e.g., by being born there) or qualify to vote in
city or town elections by being a citizen residing in the town or city.
Personal rights defined by a certain qualifying role may not be bought or sold
since the buyer may not have the qualifying role and would not need to buy the
rights if they had the qualifications. By the same token, personal rights may
not be bequeathed or inherited; the rights die with the person. Voting rights
that are held as personal rights are always one-person/one-vote since either
the person qualifies or doesn't; there are no multiple qualifications.

A \textit{property right} is a right that a person may have independent of any
qualifications and in any quantity. Property rights are, in general,
transferable by purchase and sale, by gift, or by inheritance. The voting
rights attached to common shares in a conventional corporation are property
rights that may be held in any quantity.

How is residual claimancy determined in different business organizations? In a
cooperative corporation, residual claimancy is a personal right based on the
qualifying role called ``patronage'' in the
cooperative, e.g., working in a worker cooperative, purchasing in a consumer
cooperative, banking in a credit union, and so forth. When residual claimancy
is a personal right instead of a property right, then those persons are not
properly called ``owners'' since their rights
do not have characteristics of property rights. They are typically called
``members'' although shareholders in a
conventional corporation are sometimes called \textquotedblleft
members,'' e.g., ``the terms `shareholders'
and `members' may be used interchangeably'' \cite[p.
304]{harrigan:UKcorplaw}. One way to conceptualize the conventional
corporation is to think of it as a ``cooperative'' where there is no patronage requirement, the
zero-patronage cooperative, so the residual claimancy rights become
free-floating property rights.

A share in a partnership is quite unlike a corporate share; it is not a
free-floating property right that may be bought and sold. A share in a
partnership is more akin to a personal right since a partner needs to qualify
by working in the partnership and being accepted by the partners--although
there may be some hybrid characteristics such as unequal voting powers.

\section{Why `fair market valuations' do not apply to cooperatives, ESOPs, or
partnerships}

\subsection{ESOPs and worker cooperatives}

Membership in a worker cooperative as well as other cooperatives is a personal
right based on patronage in the cooperative. Employee Stock Ownership Plans
(ESOPs) are sometimes billed as ``turning workers into
capitalists'' \cite{kelso-Hetter:capitalists} but the actual
structure of the ESOP transaction is quite different. In the U.S. ESOP, there
is a special type of pension trust associated with a company. All the
employees of the company are automatically qualified to be members or
beneficiaries of the trust \textit{based on their qualifying role of
employment} (usually beyond a probationary period) in the company. And being a
member of the trust means the employees have an internal capital account (ICA)
with a certain number of shares attributed to them. When the company makes a
tax-deductible ESOP contribution to the ESOP to pay for the shares sold by
shareholders (e.g., a retiring owner), then the paid-for shares are allocated
to the members' ICA based on their labor as measured by their pay. Hence being
a member in the ESOP and the distribution of the value of ESOP contributions
within the ESOP are both based on the person's qualifying role as an employee
in the company. Therefore those are personal rights and, accordingly, the shares
``in'' an employee's ICA cannot be sold,
gifted, mortgaged, or bequeathed. And when an employee exits the qualifying
role of working in the company, then their membership in the ESOP is also
terminated. These are all the characteristics of personal rights.

The ESOP way for employees to get (indirect) ownership of company shares
should be contrasted with the Employee Share Purchase Plans (ESPPs) where the
employees buy shares from the company or other shareholders by paying for them
out of payroll (often with discounts on company-supplied shares or company
contributions). Aside from the discounts, that is basically an ordinary
exchange of property rights like any other purchase and sale of company shares
by outsiders.

Worker cooperatives are an even simpler example of workers getting membership
(i.e., voting and residual claimancy rights) as personal rights based on their
qualifying role of patronage, i.e., working in the cooperative (usually beyond
a probationary period).

In both ESOPs and worker cooperatives, the people working in the company get
membership including their residual claimants status as personal rights based
on them currently working in the company.

Hence in the application of the standard ``fair market valuation'' of the ESOP shares or worker cooperative
memberships, the current members (qua \textit{current} members) are not the
residual claimants in the future time periods since the residual claimancy is
then held by the \textit{future} worker-members in the enterprise. In the
formulas of our simple model, the future net cashflow $PQ-VC-WL$ is not
assigned to the current members but to the future members of those firms in
those future time periods. In other words, the future members are not the
employees (with the capped claims $WL$) of the current members. Hence the
valuation of the shares in an ESOP by the standard discounted cashflow method
is inappropriate. And, of course, the same holds if the standard valuation
methods were (mis)applied to value ``membership shares'' in a worker cooperative.

Since the standard valuation methods produce the value $V=NAV+Goodwill$, our
analysis shows that the future profits, with the discounted value of
$V_{0}=Goodwill$, will be claimed in an ESOP or worker cooperative by the
future members, so the net value held by the current members is the $NAV$, the
net asset value as shown on the company's balance sheet. 

When an ESOP is being
first set up, the seller of shares in a non-public company may want a market
valuation since they have the alternative of selling to other buyers. But once
shares are inside the one-way trapdoor of the ESOP, those indirect owners get
their shares as personal rights based on the role of being an employee in the
underlying company, so the usual `fair market valuation' is inappropriate.
However, the current law for U.S. ESOPs in privately-held companies requires
an expensive annual valuation by standard methods. This regulation is not only
expensive but inappropriate. No additional valuation or expense is required to
extract the $NAV$ from the company's year-end balance sheet. For an exiting or
retiring employee, the company is legally required to repurchase the shares in
the employee's ICA with this inflated and inappropriate market valuation. This
seems to be one of the contributing factors to the number of ESOP sellouts
necessary to meet their repurchase liability.

When the company makes an ESOP contribution to the ESOP to be passed through
to pay off the credits to buy the shares, the principal portion of the loan
payments is distributed between the member ICAs in accordance with their
labor, not the number of shares already in the ICAs. But when the shares are
revalued, either according to the external market valuation or the increase in
$NAV$, then that value is distributed between the share-denominated ICAs in
accordance with the number of shares in the accounts, not according to the
labor (however measured) of the account-holders.

This raises the question of
share-denominated ICAs as in the U.S. ESOPs or value-denominated ICAs as in Mondragon-type worker cooperatives. In the Employee Ownership Trusts (EOTs) in the United Kingdom, there are no individual capital accounts, only collective ownership, so the question is moot. In Canada, legislation was passed for EOTs which follows the UK model except that it allows EOTs with ICAs. The first ESOP legislation (with mandatory ICAs) outside of the United States was recently (October 2025) passed in Slovenia. In that modified ESOP model, called the Coop-ESOP
\cite{ell-gonza-greg:coop-esop} or EurESOP, it is recommended that the ICAs be
value-denominated so the increases (or decreases) in $NAV$ each year would be
distributed between the ICAs in accordance with the labor of the members. Then
no further annual market valuations are needed--in addition to not being appropriate.

\subsection{Partnerships}

The residual claimants in a partnership are the partners. Since being a
partner is essentially a personal right, the same arguments about the market
valuation of shares in a partnership apply, \textit{mutatis mutandis}, to
partnerships. Moreover, partnerships have ICAs which are called
``partner capital accounts'' or ``equity accounts.'' The name
``equity accounts'' is a bit of a misnomer
since a party with more equity would have more votes and a larger share of the
residual, but neither voting power nor shares in the net income are related to
the size of the Partner Capital Accounts (PCAs).\footnote{See any accounting
text with a chapter on partnership accounting such as \cite{warren:accting} or
\cite{wild-shaw:accting}, or simply do an internet search on ``partner capital accounts.''} The PCAs, like the ICAs in a
worker cooperative, are really a subordinate form of debt no matter where
accounting practice places them on the balance sheet.

When a new partner enters a partnership, they will typically make a capital
contribution. The net value of the contribution (which could be a
proprietorship with both assets and liabilities folded into the partnership),
is added to the new partner's PCA. Each year the partner's share in the net
income (as specified in the partnership agreement) is added to their PCA and
any cash withdrawals are subtracted from the PCA.

When a partner retires or otherwise exits the partnership, then the terminal
value in their PCA is the amount owed to them (usually with interest).
Problems arise, however, when share-based reasoning enters into the matter. An
exiting partner may think of their share in the partnership as being like a
corporate share or the partnership may be legally structured as some variation
on a limited liability corporation with shares. The exiting partner may argue
that the ``name'' (i.e., reputation) was
little-known when they entered but is now a well-known entity so they should
be paid for this increase in ``goodwill.'' But the future partners, who would be paying the exiting partner for this
``goodwill,'' as in the standard market
valuation, are residual claimants in their own right and thus do not need to
buy those rights from the current partner (including the exiting one). Again,
it is thus inappropriate to treat the future partners as having only the
capped returns (e.g., $WL$) as if the future residual rights were owned by the
current partners and as is assumed in the standard cashflow `fair market' valuations.

\begin{quotation}
\noindent In such circumstances it is relatively easy for partners to move in
and out of the ranks. The ``naked out'' part
of this means that a retiring partner does not realise a capital return for
goodwill built up over time. What the partner gets is a share of profits over
the time he or she was a partner. This business model can stand the test of
time. There are no distracting discussions on what ``my
equity'' is worth and ``how will I get paid
that increase in value\textquotedblright? \cite{fieldfisher:2015}
\end{quotation}

\noindent Thus the exiting partner goes out with the payout of their PCA
representing the retained (i.e., not withdrawn) ``share of
profits over the time he or she was a partner'' but is
``naked out'' concerning any additional
payment for ``goodwill built up over time.''

\section{Remarks on some related issues}

The analysis given here applies to shares in an ESOP and to partnership
shares. But prior to the establishment of an ESOP, there is the question of
how the seller of shares potentially to the ESOP might value their shares. The
analysis does not apply to that question since the seller has the option of
selling in a market transaction where the buyer would be treating the future
workers in the enterprise as just employees so a valuation method
incorporating that assumption would not be inappropriate. 

What should an ESOP law in the U.S. or elsewhere say? The analysis implies
that the shares in an ESOP should be marked to Net Asset Value (NAV) as
specified on the balance sheet of the associated company. The acquired shares
are an asset of the ESOP so if they were purchased at a premium over NAV, then
they should be marked down to NAV. The Individual Capital Accounts (ICAs) in
an ESOP should be viewed as a form of subordinate debt of the ESOP to the
employee-members. There should be another true equity residual account such as
a Collective Account that is not individuated. On the Liability and Equity
side of the ESOP's balance sheet, the markdown should be debited to that
residual equity account. After the ESOP is established, the analysis given
here implies that any market valuation of the shares in the ESOP is
inappropriate (in addition to being costly). It is not a matter of doing a
market valuation but then applying a non-marketable discount on the shares.
Instead, the shares should be marked to NAV which incurs no cost since the NAV
is available on the company's balance sheet. This would require a major change
in U.S. ESOP law. 

The point that the future ESOP members are not the employees of the current
ESOP members is independent of the degree of the ESOP's ownership of the
company. It is to be expected that in a profitable company (i.e., positive
goodwill), the external market valuation of the shares in the ESOP will be
higher than the internal valuation at NAV and the same holds in a partnership.
That is, it will typically be profitable to treat the future ESOP members or
future partners as being employees of the current members/partners instead of
being residual claimants in their own rights. In those circumstances, it will
always be profitable for the current members of an ESOP to terminate the ESOP
in a sellout. 

This is not a new discovery. It is only a new technical perspective on an old
factoid about ordinary human proclivities. In the old debate about why there
are not more employee-owned firms, the rather obvious answer was expressed in
the``founders' decision.'' When one or more
people found and co-own a new firm, they face the founders' decision about
whether to treat all the later people to work in the company as co-owners or
just as employees. 

There is even a political version of the fact pattern. After a revolution, the
leaders of the revolution face the decision to form a democracy where they
would at best hold only delegated power and at worst be voted out--or would
they form a new autocracy. For instance, America became a democratic republic
because George Washington resisted the efforts of his officer corps to be an
aristocracy with Washington as monarch and he followed the example of
Cincinnatus by returning to his farm \cite{wills:cinn}.

The alternative to market valuation is individual capital accounts (ICAs) in
some ESOPs and worker cooperatives or partner capital accounts (PCAs) in
partnerships. These ICAs keep track of the value that should ultimately be
returned to the account-holder. For example, in a partnership, a new partner
may initiate their PCA with a substantial contribution and then at the end of
each fiscal year, the partner's share of the profit is credited to their PCA
and their cash withdrawals are subtracted from their PCA. In the world of
worker cooperatives, the use of ICAs seems to have been pioneered by the
Mondragon cooperatives. \cite{whytes:mondragon}. In American agricultural, producer, and
marketing cooperatives, there has long been the practice of having
``written notices of patronage allocations'' for patronage dividends that were not paid out in cash \cite{zeuli:coop}. Thus one
might think of the accumulation of a member's written notices as their ICA.
But the overt introduction of ICAs in American worker cooperatives started in
the late 1970s with the Model Bylaws of the Industrial Cooperative Association
in Boston following the Mondragon example \cite{ICA:bylaws}. 

The logic of these ICAs was identical to that of partnership capital accounts.
There might be an initial (and usually nominal) ``membership
fee'' as the initial balance in a member's ICA. The residual
claim of the cooperative members was to the value added, i.e., the revenues
minus the non-labor costs. Part of the value-added is paid out in wages during
the year--which Mondragon called ``advances'' on a member's share of the value-added. At the end of each fiscal year, some
profits are retained (or are already retained by net investments during the
year) and there might be some cash distribution of profits. The division
between members is based on their ``patronage'' which is usually measured by their annual wages
or salaries. Each member's share of the \textit{retained} profit is added to their ICA.
If losses were made, then the proportionate share is subtracted from their
ICA. Thus each member shares in the \textit{realised} profits made by the
cooperative. Since there is no valuation by the conventional methods, there is
no imputation to the member ICAs of the discounted present value of
anticipated future profits--since those would be earned by the future members
of the cooperative. And since the ICAs are essentially a form of subordinate
debt to members, the ICAs should bear interest.

If we compare these\ ICAs in the Mondragon-type worker cooperatives and the
ICAs in the U.S. ESOPs updated each year with market valuations, then both
accounts go up with the realised profits that are retained in the company. The
difference is that the Mondragon-type ICAs do not incorporate any discounted
present value of expected future profits while the inappropriate market
valuations used to update the U.S. ESOP ICAs involve some such estimates. The
problem is not that the future profits are uncertain. The problem is that for
both Mondragon-type worker cooperatives and ESOPs, the future members are not
employees of the current members but are residual claimants of those profits
realised in the future.

\section{Conclusion}

The standard discounted cashflow method of valuation treats the (pure) profit
rights in future time periods as the property rights of the current
shareholders in a corporation. Of course, the size of future residual profits
is uncertain and the contractual behavior of future suppliers and customers is
not ``owned'' by the current shareholders.
But that is not our main point. Our argument is that in certain legal
organizations such as worker cooperatives, ESOPs, and partnerships, the future
profit rights are not owned by the \textit{current} residual claimants. Those
future residual rights are held as personal rights by the future members of
the cooperative, ESOP, or partnership. Hence the usual discounted cashflow
valuation methods are inappropriate and wrong for those business organizations.

\section{Appendix: The Miller-Modigliani model for valuations}

\subsection{Introduction}

In Miller and Modigliani's seminal (and Nobel Prize winning) paper
\cite{Miller-Modig:valuation}, they derived four formulas for the fair market
value of a corporation and proved that the four formulas were equivalent:

\begin{enumerate}
\item the discounted cashflow approach,

\item the current earnings plus future investment opportunities approach,

\item the stream of dividends approach, and

\item the stream of earnings approach.
\end{enumerate}

None of the approaches parsed the value into the value of the current owned
net assets plus the property the corporation expected to appropriate in the
future, i.e., the $NAV$ + goodwill approach.

Our purpose in this appendix is to prove the equivalence of that fifth
approach with the discounted cashflow approach and thus the equivalence of all
five approaches. We start with the stream of dividends approach to establish notation.

The balance sheet accounts of \textit{Gross Asset Value} $GAV(t)$, the
\textit{Debt} $D(t)$, and the \textit{Net Asset Value} $NAV(t)$ are
\textit{stock} variables that represent a value at a point in time $t$ so the
\textit{balance sheet equation} at time $t$ is:

\begin{center}
$GAV(t)=D\left(  t\right)  +NAV(t)$.
\end{center}

Other accounts such as the income statement accounts or cashflow statement
accounts are \textit{flow }variables that represent the change in the value of
a stock variable. That change in the value of a stock variable is represented
by a lower case $d$ in front of the stock variable:
$dGAV(t)=GAV(t+1)-GAV\left(  t\right)  $ and $dNAV(t)=NAV\left(  t+1\right)
-NAV\left(  t\right)  $. In our previous simplified model, we assumed that the
fixed assets or ``machine'' had a constant
market value $C$, i.e., no depreciation, and then died after $n$ years with
salvage value $S$. Miller and Modigliani (MM) make the more realistic
assumption that there is the market value $C(m)$ for a vintage
$m$ machine, i.e., a machine in use for $m$ years. Thus the
\textit{depreciation} in a year is $C(m-1)  -C(m)  $.

MM assume only simple common shares in the corporation where:

\begin{itemize}
\item $n\left(  t\right)  $ = number of shares outstanding at time $t$,

\item $v\left(  t\right)  $ = price per share at time $t$,

\item $V\left(  t\right)  =n\left(  t\right)  v\left(  t\right)  $ = value of
all outstanding shares at time $t$ = value of the corporation at time $t$,

\item $\operatorname{div}\left(  t\right)  $ = dividends per share at time
$t+1$ to shareholders at time $t$, and

\item $Div\left(  t\right)  =n(t)\operatorname{div}\left(  t\right)  $ = total
dividends paid at time $t+1$.
\end{itemize}

\subsection{The stream-of-dividends formula}

The formulas for $V\left(  t\right)  $, the value of the corporation, were
developed in the MM framework where arbitrage under conditions of perfect
competition and assumed certainty, capital must be paid the same rate of
return whether it is loaned out at interest or invested in corporate shares.

On one share at value $v\left(  t\right)  $ at time $t$, the end-of-period
return is the dividends $\operatorname{div}\left(  t\right)  $ plus the
capital gains $v\left(  t+1\right)  -v\left(  t\right)  $. If that amount of
money was loaned out at interest $r$, then the return is $rv\left(  t\right)
$. If the returns are unequal, then arbitrage under the assumed conditions
will equalize the return to yield the \textit{arbitrage equation}:

\begin{center}
$rv\left(  t\right)  =\operatorname{div}\left(  t\right)  +v\left(
t+1\right)  -v\left(  t\right)  $.
\end{center}

Thus $v\left(  t\right)  =\frac{\operatorname{div}\left(  t\right)  +v\left(
t+1\right)  }{1+r}$ so multiplying through by $n\left(  t\right)  $ yields:

\begin{center}
$V\left(  t\right)  =\frac{n\left(  t\right)  \left[  \operatorname{div}%
\left(  t\right)  +v\left(  t+1\right)  \right]  }{1+r}$.
\end{center}

To obtain the stream-of-dividends formula for $V\left(  t\right)  $, we use
the arbitrage equation to expand the $v\left(  t+1\right)  $ term to obtain:

\begin{center}
$V\left(  t\right)  =\frac{n\left(  t\right)  \operatorname{div}\left(
t\right)  }{1+r}+\frac{n\left(  t\right)  \operatorname{div}\left(
t+1\right)  }{\left(  1+r\right)  ^{2}}+\frac{n\left(  t\right)  v\left(
t+2\right)  }{\left(  1+r\right)  ^{2}}$.
\end{center}

Then repeating this use of the arbitrage equation yields the
\textit{stream-of-dividends formula}:

\begin{center}
$V\left(  t\right)  =\sum_{k=1}^{\infty}\frac{n\left(  t\right)
\operatorname{div}\left(  t+k-1\right)  }{\left(  1+r\right)  ^{k}}$.
\end{center}

\noindent Thus the value of the corporation at time $t$ is the discounted
present value of the stream of dividends ``that accrues to the
shares of record as of the start of period $t$'' \cite[p.
419]{Miller-Modig:valuation}, see also \cite[p. 87, formuila 2.19b]%
{fama-miller:finance}. The shares in a conventional corporation include the
residual claimant rights, so the additional future residual claimants, i.e.,
owners of new shares purchased as property rights after time $t$, do not add
to the value of the shares held at time $t$.

The usual cookbook valuation formulas as well as the MM formulas all assume
that the corporate assets are used as in the active case. If instead we
consider the \textit{passive} use of the corporate assets by lending out the
Gross Asset Value $GAV(t)$ at the interest rate $r$, then the gross return is
$rGAV\left(  t\right)  $ so after paying the interest on the debt $D\left(
t\right)  $, the net return is: $r\left[  GAV\left(  t\right)  -D\left(
t\right)  \right]  =rNAV\left(  t\right)  $. To compute the discounted present
value of the steady infinite stream of values $rNAV\left(  t\right)  $, we
need to prove $\sum_{t=1}^{\infty}\frac{r}{\left(  1+r\right)  ^{t}}=1$ which
is the present value of one dollar loaned out at the interest rate $r$. For
$X:=\sum_{k=1}^{\infty}\frac{r}{\left(  1+r\right)  ^{k}}$, if we multiply
through by $1+r$, then $\left(  1+r\right)  X=\sum_{k=1}^{\infty}%
\frac{r\left(  1+r\right)  }{\left(  1+r\right)  ^{k}}=\sum_{k=1}^{\infty
}\frac{r}{\left(  1+r\right)  ^{k-1}}=r+X$ so solving for $X$ yields: $rX=r$
or $X=1$. Thus we have:

\begin{center}
$\sum_{k=1}^{\infty}\frac{\not r  NAV\left(  t\right)  }{\left(  1+r\right)
^{k}}=NAV\left(  t\right)  $.
\end{center}

\noindent This is the MM version of our previous result in the simple model
that the value of the machine \textit{in the passive case} is $C$.

\subsection{The discounted cashflow formula}

Let $A\left(  t\right)  $ be the net accounting profit at time $t$ (all cash
transactions) so $A\left(  t\right)  =\mathcal{R}(t)-COGS\left(  t\right)
-Depr\left(  t\right)  $ where $\mathcal{R}\left(  t\right)  $ is the cash
revenue, $COGS\left(  t\right)  $ is the costs of goods sold (including labor
and debt interest), and $Depr\left(  t\right)  $ is the depreciation for the
time period. Hence $A\left(  t\right)  +Depr\left(  t\right)  -DIV\left(
t\right)  $ is the cash available from operations for the gross investment
$I\left(  t\right)  +Depr\left(  t\right)  $ where $I\left(  t\right)  $ is
the net investment during the time period. The rest of the planned investment
$I\left(  t\right)  +Depr\left(  t\right)  $ has to come from issuing
$m\left(  t+1\right)  $ new shares at the price $v\left(  t+1\right)  $ per
share for a total subscription $Sub\left(  t+1\right)  :=m\left(  t+1\right)
v\left(  t+1\right)  $. Hence we have the equation:

\begin{center}
$Sub\left(  t+1\right)  =I\left(  t\right)  +Depr\left(  t\right)  -\left[
A\left(  t\right)  +Depr(t)-DIV\left(  t\right)  \right]  =I\left(  t\right)
-\left[  A\left(  t\right)  -DIV\left(  t\right)  \right]  $.
\end{center}

\noindent The total value of the corporation $V\left(  t+1\right)  $ is the
total value of the old shares plus the subscription of new shares so:

\begin{center}
$V\left(  t+1\right)  =n\left(  t\right)  v\left(  t+1\right)  +Sub\left(
t+1\right)  $.
\end{center}

\noindent This can then be plugged into the previous equation and simplified:

\begin{center}
$V\left(  t\right)  =\frac{n\left(  t\right)  \left[  \operatorname{div}%
\left(  t\right)  +v\left(  t+1\right)  \right]  }{1+r}=\frac{1}{1+r}\left[
DIV\ \left(  t\right)  +V\left(  t+1\right)  -Sub\left(  t+1\right)  \right]
$

$=\frac{1}{1+r}\left[  DIV\left(  t\right)  +V\left(  t+1\right)  -\{I\left(
t\right)  -\left[  A\left(  t\right)  -DIV\left(  t\right)  \right]
\}\right]  $.
\end{center}

\noindent Then the two $DIV\left(  t\right)  $ terms cancel out--which is the
basis for MM's famous ``dividend irrelevance\textquotedblright%
\ thesis. The resulting formula is:

\begin{center}
$V\left(  t\right)  =\frac{1}{1+r}\left[  A\left(  t\right)  -I\left(
t\right)  +V\left(  t+1\right)  \right]  $.
\end{center}

\noindent Then by making the repeated substitutions for the $V(t+1)$, we have
the formula:

\begin{center}
$V\left(  t\right)  =\sum_{k=1}^{\infty}\frac{1}{\left(  1+r\right)  ^{k}%
}\left[  A\left(  t+k-1\right)  -I\left(  t+k-1\right)  \right]  $.
\end{center}

\noindent The \textit{cash receipts} $\mathcal{R}\left(  t\right)  $ are the
net accounting profits $A\left(  t\right)  $ plus the depreciation
$Depr\left(  t\right)  $ and costs of goods sold $COGS(t)$ so: $\mathcal{R}%
\left(  t\right)  =A\left(  t\right)  +Depr\left(  t\right)  +COGS\left(
t\right)  $. The \textit{cash outlays} are $\mathcal{O}\left(  t\right)
=I\left(  t\right)  +Depr\left(  t\right)  +COGS\left(  t\right)  $ so the
$Depr\left(  t\right)  +COGS\left(  t\right)  $ terms cancel and we have the
\textit{discounted cashflow formula}:

\begin{center}
$V\left(  t\right)  =\sum_{k=1}^{\infty}\frac{1}{\left(  1+r\right)  ^{k}%
}\left[  \mathcal{R}\left(  t+k-1\right)  -\mathcal{O}\left(  t+k-1\right)
\right]  $.
\end{center}

\subsection{The NAV + goodwill formula}

The net asset value $NAV\left(  t+1\right)  $ is the previous net asset value
$NAV\left(  T\right)  $ minus the depreciation $Depr\left(  t\right)  $ plus
the gross investment $I\left(  t\right)  +Depr\left(  t\right)  $ so the
depreciation terms cancel in $dNAV\left(  t\right)  =NAV\left(  t+1\right)
-NAV\left(  t\right)  =I\left(  t\right)  .$ Hence substituting into $V\left(
t\right)  =\sum_{k=1}^{\infty}\frac{1}{\left(  1+r\right)  ^{k}}\left[
A\left(  t+k-1\right)  -I\left(  t+k-1\right)  \right]  $, we have the formula:

\begin{center}
$V\left(  t\right)  =\sum_{k=1}^{\infty}\frac{1}{\left(  1+r\right)  ^{k}%
}\left[  A\left(  t+k-1\right)  -dNAV\left(  t+k-1\right)  \right]  $.
\end{center}

The \textit{economic or pure profit} $\pi\left(  t+k-1\right)  $ is the
accounting profit $A\left(  t+k-1\right)  $ minus the implicit interest cost
of tying up the capital $NAV\left(  t+k-1\right)  $ for the time period so:

\begin{center}
$\pi\left(  t+k-1\right)  :=A\left(  t+k-1\right)  -rNAV\left(  t+k-1\right)
$.
\end{center}

The \textit{goodwill} $GW\left(  t\right)  $ is defined as the discounted
present value of the future pure profit so the goodwill at time $t$ is:

\begin{center}
$GW\left(  t\right)  :=\sum_{k=1}^{\infty}\frac{1}{\left(  1+r\right)  ^{k}%
}\pi\left(  t+k-1\right)  =\sum_{k=1}^{\infty}\frac{1}{\left(  1+r\right)
^{k}}\left[  A\left(  t+k-1\right)  -rNAV\left(  t+k-1\right)  \right]  $.
\end{center}

\noindent Then

\begin{center}
$NAV\left(  t+k-1\right)  =NAV\left(  t\right)  +dNAV(t)+dNAV\left(
t+1\right)  +...+dNAV\left(  t+k-2\right)  =NAV\left(  t\right)  +\sum
_{j=0}^{k-2}dNAV\left(  t+j\right)  $
\end{center}

\noindent where the sum $\sum_{j=0}^{k-2}dNAV\left(  t+j\right)  $ only begins
for $k=2,3,...$ so substituting into the formula for $\pi\left(  t+k-1\right)
$ gives:

\begin{center}
$\pi\left(  t+k-1\right)  =A\left(  t+k-1\right)  -rNAV\left(  t\right)
-r\sum_{j=0}^{k-2}dNAV\left(  t+j\right)  $ for $k=2,3,...$.
\end{center}

Using the identity $\sum_{k=1}^{\infty}\frac{r}{\left(  1+r\right)  ^{k}}=1$,
$\sum_{k=1}^{\infty}\frac{1}{\left(  1+r\right)  ^{k}}rNAV\left(  t\right)
=NAV\left(  t\right)  $ in sum for $GW\left(  t\right)  $. The tricky part is
the double-sum term in $GW\left(  t\right)  $:

$\sum_{k=2}^{\infty}\frac{1}{\left(  1+r\right)  ^{k}}r\sum_{j=0}%
^{k-2}dNAV\left(  t+j\right)  $. The way to attack it is to pick out all the
occurrences of $dNAV\left(  t\right)  $, $dNAV\left(  t+1\right)  $,.... For
instance, the $dNAV\left(  t\right)  $ occurs with $j=0$ for $k=2,...$so the
sum of those terms is:

\begin{center}
$\sum_{k=2}^{\infty}\frac{r}{(1+r)^{k}}dNAV(t)=\sum_{k=1}^{\infty}\frac
{r}{\left(  1+r\right)  ^{k}}\frac{dNAV\left(  t\right)  }{\left(  1+r\right)
}=\frac{dNAV\left(  t\right)  }{\left(  1+r\right)  }$.
\end{center}

\noindent Similarly, the term $dNAV\left(  t+1\right)  $ occurs with $j=1$ for
$k=3,...$ so the sum of those terms is:

\begin{center}
$\sum_{k=3}^{\infty}\frac{r}{(1+r)^{k}}dNAV(t+1)=\sum_{k=1}^{\infty}\frac
{r}{\left(  1+r\right)  ^{k}}\frac{dNAV\left(  t+1\right)  }{\left(
1+r\right)  ^{2}}=\frac{dNAV\left(  t+1\right)  }{\left(  1+r\right)  ^{2}}$.
\end{center}

\noindent The general term $dNAV\left(  t+T-1\right)  $ occurs with $j=T-1$
for general $k=T+1,...$ so the sum of those terms is:

\begin{center}
$\sum_{k=T+1}^{\infty}\frac{r}{(1+r)^{k}}dNAV(t+T-1)=\sum_{k=1}^{\infty}%
\frac{r}{\left(  1+r\right)  ^{k}}\frac{dNAV\left(  t+T-1\right)  }{\left(
1+r\right)  ^{T}}=\frac{dNAV\left(  t+T-1\right)  }{\left(  1+r\right)  ^{T}}$.
\end{center}

Finally, the double sum in $GW\left(  t\right)  $ is:

\begin{center}
$\sum_{k=2}^{\infty}\frac{1}{\left(  1+r\right)  ^{k}}r\sum_{j=0}%
^{k-2}dNAV\left(  t+j\right)  =\sum_{k=1}^{\infty}\frac{dNAV(t+k-1)}{\left(
1+r\right)  ^{k}}=\sum_{k=1}^{\infty}\frac{I\left(  t+k-1\right)  }{\left(
1+r\right)  ^{k}}$.
\end{center}

Collecting terms, we have:

\begin{center}
$GW\left(  t\right)  =\sum_{k=1}^{\infty}\frac{A\left(  t+k-1\right)
}{\left(  1+r\right)  ^{k}}-NAV\left(  t\right)  -\sum_{k=1}^{\infty}%
\frac{dNAV(t+k-1)}{\left(  1+r\right)  ^{k}}$
\end{center}

\noindent so moving the $NAV\left(  t\right)  $ term to the other side, we
have the \textit{NAV + Goodwill formula\footnote{This proof was first
published in Chapter 12 of \cite{ell:EAPT}.}} for the value of the corporation
at time $t$:

\begin{center}
$NAV\left(  t\right)  +GW\left(  t\right)  =\sum_{k=1}^{\infty}\frac
{1}{\left(  1+r\right)  ^{k}}\left[  A\left(  t+k-1\right)  -I\left(
t+k-1\right)  )\right]  $

$=\sum_{k=1}^{\infty}\frac{1}{\left(  1+r\right)  ^{k}}\left[  \mathcal{R}%
\left(  t+k-1\right)  -\mathcal{O}\left(  t+k-1\right)  \right]  =V\left(
t\right)  $.
\end{center}

MM give four equivalent formulas for $V\left(  t\right)  $ and this is a fifth
equivalent formula (see also \cite{edey:superprofits} where the pure profits
are called ``super profits''). Moreover, it is
the only formula that parses the property rights into the actual net property
rights at time $t$, namely $NAV\left(  t\right)  $, plus the discounted
present value of the expected property rights appropriated in the active case
in the future time periods. MM assume in their idealized model that the future
values are known with certainty (so they don't have to consider risk-adjusted
discount rates), but such an assumption does not change the legal fact that
the present corporation at $t$ does not have legally enforceable contracts
with all future suppliers or customers, i.e., no present property rights to
future \textit{pure} profits. The future claims from the net property owned by
the corporation at prior times is recognized by the $rNAV\left(  t+k-1\right)
$ term subtracted from the accounting profits $A\left(  t+k-1\right)  $ to get
the pure profits $\pi\left(  t+k-1\right)  $.

Our main point is that the shareholders at time $t$ do not own the future pure
profits that will accrue to the residual claimants at all those future time periods.

In spite of the complications in the proof of the NAV(t) + Goodwill formula,
there is a certain logic to the formula. The value $NAV\left(  t\right)  $ is
owned by the shareholders at time $t$. In each future time period, the net
accounting profit $A\left(  t+k-1\right)  $ is the value created during that
time period after subtracting explicit expenses and depreciation but it does
not subtract the implicit cost $rNAV\left(  t+k-1\right)  $ of tying up that
capital for the time period so it is really the pure or economic profit
$\pi\left(  t+k-1\right)  =A\left(  t+k-1\right)  -rNAV\left(  t+k-1\right)  $
that represents the \textit{new} value created in the time period. The
discounted present value of those new values is the goodwill so the total
value creation in the active case is $NAV\left(  t\right)  +GW\left(
t\right)  =V\left(  t\right)  $. In an ESOP, worker cooperative, or
partnership, those future new values accrue to the then future residual
claimants so the proper value accruing to the current residual claimants is
$NAV\left(  t\right)  $. Hence the standard `fair market valuation,' obtained
either using shorthand cookbook formulas or using the full MM framework
formulas is inappropriate for those types of firms.


\begin{thebibliography}{99}                                                                                               %


\bibitem {edey:superprofits}Edey, H. C. 1962. ``Business
Valuation, Goodwill and the Super-Profit Method.'' In
\textit{Studies in Accounting Theory}, edited by W. T Baxter and S. Davidson. Irwin.

\bibitem {ell:ownership-myth}Ellerman, David. 1975. ``The
`Ownership of the Firm' Is a Myth.'' In
\textit{Organizational Democracy: Participation and Self-Management}, edited
by David Garson and Michael P. Smith. Sage Publications.

\bibitem {ell:EAPT}Ellerman, David. 1982. \textit{Economics, Accounting, and
Property Theory}. D.C. Heath.

\bibitem {ell:goodwill}Ellerman, David. 2008. ``Goodwill: A
Present Property Right or Only An Anticipated Future Right?'' \textit{FSR Forum (Financial Studies Association Rotterdam)} August 2008: 23--25.

\bibitem {ell:juris-econ}Ellerman, David. 2021. \textit{Putting Jurisprudence
Back into Economics: On What Is Really Wrong with Today's Neoclassical
Theory}. SpringerNature. https://doi.org/10.1007/978-3-030-76096-0.

\bibitem {ell-gonza-greg:coop-esop}Ellerman, David, Tej Gonza, and Gregor
Berkopec. 2022. ``European Employee Stock Ownership Plan
(ESOP): The Main Structural Features and Pilot Implementation in
Slovenia.'' \textit{SN Business \& Economics} 2 (12): 186. https://doi.org/10.1007/s43546-022-00363-7.

\bibitem {ell-gonza:coops}Ellerman, David, and Tej Gonza. 2024.
``Worker Cooperatives and Other `Cooperatives.'
'' In \textit{Routledge Handbook on Cooperative Economics and
Management}, edited by Jerome Warren, Jamin H\"{u}bner, Lucio Biggiero, and
Kemi Ogunyemi. Routledge. https://doi.org/10.4324/9781003449850-8.

\bibitem {fama-miller:finance}Fama, Eugene F., and Merton H. Miller. 1972.
\textit{The Theory of Finance}. Holt, Rinehart and Winston.

\bibitem {fieldfisher:2015}Fieldfisher. 2015. ``Employee Buy
Outs: An Alternative Succession Solution for Professional
Partnerships.'' \textit{Insights, November 11}. https://www.fieldfisher.com/en/insights/employee-buy-outs-an-alternative-succession-solution-for-professional-partnerships.

\bibitem {friedman-ordway:appraisal}Friedman, Jack P., and Nicholas Ordway.
1988. \textit{Income Property Appraisal and Analysis}. Prentice Hall. https://doi.org/10.1007/978-3-030-76096-0.

\bibitem {harrigan:UKcorplaw}Hannigan, Brenda. 2012. \textit{Company Law 3rd
Ed}. Oxford University Press.

\bibitem{ICA:bylaws}Industrial Cooperative Association. 1983. \textit{ICA MODEL BY-LAWS FOR A WORKER COOPERATIVE Version II}. Somerville MA.

\bibitem {kelso-Hetter:capitalists}Kelso, Louis, and Patricia Hetter. 1967.
\textit{How to Turn Eighty Million Workers Into Capitalists on Borrowed
Money}. Random House.

\bibitem {keynes:GeneralTheory}Keynes, John Maynard. 1936. \textit{The General
Theory of Employment, Interest, and Money}. Harcourt, Brace \& World

\bibitem {Miller-Modig:valuation}Miller, Merton H., and Franco Modigliani.
1961. ``Dividend Policy, Growth, and the Valuation of
Shares.'' \textit{The Journal of Business} 34 (October 1961): 411--433.

\bibitem {robe:firm}Rob\'{e}, Jean-Philippe. 2011. ``The Legal
Structure of the Firm.'' \textit{Accounting, Economics, and
Law} 1 (1): Article 5. https://doi.org/10.2202/2152-2820.1001.

\bibitem {warren:accting}Warren, Carl S., James M. Reeve, and Jonathan E.
Duchac. 2007. \textit{Accounting 22E}. Thomson South-Western.

\bibitem {whytes:mondragon}Whyte, William Foote, and Kathleen King Whyte.
1991. \textit{Making Mondragon}. 2nd revised. ILR Press.

\bibitem {wild-shaw:accting}Wild, John J., and Ken W. Shaw. 2019.
\textit{Fundamental Accounting Principles 24th Ed}. McGraw Hill Education.

\bibitem{wills:cinn}Wills, Garry. 1984. \textit{Cincinnatus: George Washington and the Enlightenment}. Doubleday.

\bibitem{zeuli:coop} Zeuli, Kimberly A., and Robert Cropp. 2004. \textit{Cooperatives: Principles and Practices in the 21st Century A1457}. University of Wisconsin Extension--Madison. http://socialeconomyaz.org/wp-content/uploads/2011/06/Zeuli-Cropp-Cooperatives.pdf.

\end{thebibliography}
\end{document}